# «WILD CABLES» IN FUSION PLASMAS AND THEIR DIAGNOSTICS WITH SPECTRAL LINE STARK BROADENING

A.V. Demura, V.I. Kogan, A.B. Kukushkin, V.A. Rantsev-Kartinov

*INF RRC "Kurchatov Institute", Moscow 123182 Russia*

We discuss the opportunities of diagnosing the parameters of hypothetical «wild cables» in fusion plasmas via analyzing the Stark broadening of hydrogen spectral lines. The «wild cables» concept have suggested the observed long-lived skeletal structures in tokamaks and Z-pinches to be caused by a microdust-assembled skeleton. We present the results of calculating the widths of vacuum channels produced by the pressure of on-skeleton high-frequency (HF) electromagnetic waves of the TEM type, which thus might protect the skeletons from ambient high-temperature plasma. The values of amplitude of such wave appear to be compatible with the measurements of those in-plasma HF electric fields in tokamak T-10 and a gaseous Z-pinch which give observable Stark broadening of hydrogen spectral lines.

1.  **INTRODUCTION**

The phenomenon of skeletal structures [1,2] addressed in this paper is a new topic in the realm of plasma physics. This topic is being developed on the basis of the widely recognized physics of the filaments in laboratory and cosmic plasmas. This physics takes its origin from pioneering works by I. Kvartskhava and co-workers [3] and W. Bostick and co-workers [4] -- for laboratory discharges; and H. Alfvén [5], and his co-workers and followers -- for space observations (see, e.g., special survey issues [6]). The interpretation of filaments, roughly speaking, may be reduced to a sort of plasma instability, either hydrodynamic one (which results in the formation of fluid filaments of plasma density and/or temperature, electric current, etc.) or more intricate instability which should be described in terms of kinetics of particles and electromagnetic fields. At any rate, all conventional approaches to filamentation phenomena are based on the assumption that *classical* electrodynamics is sufficient for describing the *long-range* correlations/bonds in plasmas so that the recourse to quantum physics is needed only for some elementary radiative-collisional processes, like e.g. radiation emission in atomic spectral lines. This generic picture is bale to treat the ensembles of individual filaments and networking of such filaments -- first of all, short-lived, chaotic ones. Within similar frame, the phenomenon of the percolatory networking of *non-chaotic, long-lived* filaments (LLFs) has been suggested on the basis of data from Z-pinch [7,8(b)], plasma focus [9], and tokamaks [8(c)]. It was also suggested [7,8] that it is necessary to consider the long-lived filaments as a separate component of plasma -- the complimentary one to conventional, «fluid» component.

However, the view on the filaments may go far beyond the above-mentioned conventional frames. It was suggested [1,2] that only the quantum picture may explain the observed the long-range bonds inside the filaments. Within such a general quantum frame, a specific model was suggested that an LLF may possess a microsolid skeleton which might be assembled during electric breakdown, well before major plasma's birth, from wildly produced carbon nanotubes or similar nanostructures of other chemical elements.

The results of the subsequent proof-of-concept studies, aimed at verification of the above hypothesis (see e.g. [10-12]), were based on an analysis of a number of databases. This included:

(i) databases which covered a broad range of experimental conditions in various fusion plasmas (Z-pinch, plasma focus, tokamaks TM-2, T-4, T-6, T-10) and allowed, thanks to diagnostic opportunities, the resolution of the fine structure of LLFs (in particular, tubularity of straight blocks of LLFs, «networking» of straight blocks which builds up a skeleton, etc. [13(a,b,d)]) at the statistically representative level;

(ii) electron micrography of various types of dust deposit in tokamak T-10 [14];

(iii) laser shadowgraphy of initial stage of a vacuum spark discharge (electric current is less than 20% of its maximum, the plasma's self-emission is not yet detectable by the high-sensitivity detectors) [15];

(iv) high-resolution visible light imaging of plasma at electric breakdown stage of discharge in former experiments in plasma focus LV-2 and tokamak T-6 (~100 ns and ~300 mcs, respectively, before detection of discharge electric current by the Rogovsky coil) [10,16].

The following three arguments allowed [10,11] to draw a bridge between the skeletons in dust deposits and the skeletal structures in plasmas, namely:

(a) presence of tubular and cartwheel-like structures in the ranges ~10 nm to ~10 μm (in dust deposits [14]) and ~100 μm to ~10 cm (in the plasma images [13(a,b,d),15,16]),

(b) topological identity of these structures,

(c) observed trend of assembling bigger tubules from smaller ones (i.e. the self-similarity).

In this paper we discuss the opportunities of diagnosing the parameters of hypothetical «wild cables» in fusion plasmas via analyzing the Stark broadening of hydrogen spectral lines. To our mind, the present status of the concept of [1,2], as outlined above, makes it reasonable to start elaborating the details of fine diagnostics like that based on resolving the Stark broadening effects in spectral lines of plasma radiation emission.

We start (Sec. 2) with recalling the main points of the «wild cables» concept which has been suggested [13(a,c,e)] for resolving the major difficulty in the concept [1,2], namely the survivability of a condensed matter (i.e. skeletons) in a *hot* ambient plasma, of temperatures up to those in kiloelectronvolt range. The microsolid skeletons were suggested to be self-protected from an ambient high-temperature plasma by a thin vacuum channels sustained self-consistently around the skeletons by the pressure of high-frequency (HF) electromagnetic waves, thanks to the skeleton-induced conversion of a small part of the incoming «static» magnetic field (poloidal, in tokamaks, or azimuthal, in Z-pinches) into HF waves of the TEM type (a «wild cable» model [13(a,c)]). Thus, the wild cable model has proposed common, and mutually interrelated, qualitative solution to the two problems: namely, (i) survivability of skeletons, and (ii) phenomenon of nonlocal (non-diffusive, in particular, ballistic) transport of energy (the latter was reliably observed in last decade in various tokamaks, see e.g. the survey [17]).

Here, we present the results of calculating the widths of vacuum channels produced by the pressure of on-skeleton high-frequency (HF) electromagnetic waves of the TEM type, which thus might protect the skeletons from ambient high-temperature plasma (Sec. 2). The values of amplitude of such wave appear to be compatible with the measurements of those in-plasma HF electric fields in tokamak T-10 [18,19] and a gaseous Z-pinch [21] which give observable Stark broadening of hydrogen spectral lines. The details of analyzing the measurements [18,19,21] of Stark broadening of hydrogen spectral lines is given in Sec. 3.

## 2. ABOUT THE CONCEPT OF «WILD CABLES»

At this point, the concept of wild cables [13(a,c)] treats quantitatively only the quasi-stationary stage of wild cable network. However, the hypothetical life story has been suggested as well. This included a qualitative picture of electrical breakdown with allowing, in particular, for the following issues: (i) probable role of the carbon (or other chemical elements capable of forming the nanotubular structures) in assembling the macroscopic skeletons of future straight sections of observable long-lived filaments in plasma, (ii) conversion of a part of the incoming poloidal magnetic field into high-frequency TEM waves ($\omega<\omega_{pe}$), including the mode conversion in the presence of the above skeletons. Here, we recall major points of what the «wild cable» is.

The wild cable consists of three elements:
(1) a skeleton which plays a role of an inner rod (this rod serves as a guiding system for the high-frequency (HF) electromagnetic (EM) waves propagating along this rod);
(2) vacuum channel around skeleton which serves as an insulator of the cable;
(3) ambient plasma which serves as a outer screening conductor in a cylindrical guiding system for HF EM waves propagating along inner rod.

The possible scenario of wild cable operation looks as follows.

A HF valve at the plasma column surface (at the node of the nearly-standing wave) cuts the field lines of the incoming magnetic field (poloidal/azimuthal in tokamaks/Z-pinches) and, thus, forms the magnetic wave of $H_{11}$ type which propagates in the vacuum cavity around straight sections (of observed length $L_c$) of the skeleton. For the frequency $\omega_{cab}$ of this wave, one has the following relation which gives an estimate for $\omega_{cab}$ and a condition for the quasi-stationarity of the plasma system:

$$\omega_{cab} \approx (\pi c / L_{cab}) < \omega_{pe}. \qquad (1)$$

A distinct bump in the spectrum of the EM field measured [20] outside plasma column in tokamak T-10 (of the wavelength $\lambda$ of several centimeters) coincides with prediction of Eq. (1) for straight filaments seen [13(a,b)] in the far periphery of plasma column T-10. The spectra of the HF EM field in the GHz frequency range were measured in [20] in the gap between the plasma column and the chamber. They revealed a distinct bump at $\nu_C \sim (4\text{-}5)\,10^9$ Hz, of the width $\sim 2\,10^9$ Hz, which always exists in ohmic heating regimes and increases with electron cyclotron heating (this bump is a stable formation and it moves to the lower frequencies and turns into a strong peak only under condition of strong instabilities, especially disruption instability). This gives $L_C \approx 3$ cm, in reasonable agreement with the visible light data from T-10, where $L_C \sim 4\text{-}5$ cm.

Under condition of Eq. (1), the $H_{11}$ wave is effectively trapped in the cavity and, due to wiring of magnetic field lines around the skeleton, is converted, at least partly, into the wave of TEM type. The latter wave possesses a HF azimuthal magnetic field and, respectively, a HF radial electric field with amplitude $E_r(r) \sim U_0 / r$, where r is the radial coordinate in a circular cylindrical cable and $U_0$ is an effective voltage bias between the skeleton and the plasma. The radial field produces a force, the Miller's force of the type -grad($\Psi$) where $\Psi$ is the kinetic energy of charged particle oscillation in the field of HF EM wave. It is this force that is capable of self-consistently sustaining the vacuum channel around skeleton and thus protecting it from plasma particle's access. (It is assumed also that the presence of a strong static external magnetic field, like in a tokamak, doesn't influence substantially the form of the cavity, even when $\omega_{Cab} \ll \omega_{Be}$, $\omega_{Be}$ is electron gyrofrequency: the latter requires the amplitude **E** of the HF electric field to have a non-zero component parallel to external magnetic field.)

The distribution of plasma density around the inner wire can be described by a set of equations for the two-temperature quasi-hydrodynamics of plasma in a HF EM field [22,23]. Fortunately, prior to direct numeric solving equations [22,23] (whose results are shown below in Figs. 1-4), it is possible to evaluate the typical values of the parameters of the problem within a simplified approach.

Under condition $l_E \gg r_D$, where $l_E$ is the characteristic length of spatial profile of $E_r(r)$ and $r_D$ is Debye radius, one can neglect the deviation from the quasi-neutrality and arrive at a quasi-Boltzmann distribution, see [23]:

$$n_e = n_{e0} \exp(-\frac{\Psi}{T_e + T_i}), \quad \Psi = \frac{e^2 E_r^2}{4 m_e \omega_{cab}^2}, \qquad (2)$$

where $n_{e0}$ is background density of plasma electrons, and $T_e$ and $T_i$ stand for electron and ion temperatures, respectively. Equation (2) obviously gives a condition for the detachment of electrons, which limits the value of $U_0$ from the side of low values. Equation (2) is to be coupled to the condition of applicability of the concept of the Miller force, $\rho \ll l_E$ (where $\rho$ is the amplitude of electron's oscillations in the HF electric field), which limits the $U_0$ from the side of high values.

Also, the HF electric field in the cables may be related to the observable electric fields. Indeed, the wild cables are the strong sources of electrostatic oscillations in plasma, first of all, along strong external magnetic field. For the EM waves inside vacuum channels of the wild cable to be compatible with the existence of strong nonlinear waves in the ambient plasma, one may consider the cable's cavity as a soliton with such a strong reduction of the eigenfrequency (a redshift) that the soliton's velocity becomes independent from dispersion. For $W/nT < 1$, where $W = E_0^2/16\pi$, this gives the following rough estimate $(W/nT) \sim \{1 - (\omega_c/\omega_{pe})\}$.

And finally, at the quasi-stationary stage of discharge, one may evaluate the spatial distribution of the amplitude $E_{pl}$ of the electric field in plasma (not inside the wild cable!), regardless of its spectral distribution, by extrapolating the scaling law of the amplitude of the TEM wave onto ambient plasma. For the contribution of a single cable to electric field directed radially with respect to the cable's axis, one has:

$$E_{pl}(r) \sim (U_0/r). \qquad (3)$$

The above-mentioned conditions allow identify the domain of possible values of $U_0$. For instance, for tokamak case one has the following estimates [13(c)]. For $r_{Cab} \sim 1\text{-}2$ mm and comparable values of minimal and maximal values of the vacuum channel, i.e. $(r_{Cab} - r_{rod}) \sim r_{Cab}$, and $\langle r \rangle \sim 1\text{-}3$ cm ($\langle r \rangle$ is the average distance between observed individual cables in the region of observation), one has $15 < U_0(\text{kV}) < 50$ with, respectively, $E_r(r_{Cab}) > 100$ kV/cm.

An analysis of observations of Stark broadening of deuterium spectral lines (and their polarization state) at the periphery of the T-10 tokamak in the region of $T_e \sim 100$ eV, allowed [18,19] to estimate the spectral range of HF electric fields ($\omega \approx \omega_{pe} \sim 10^{11}$ Hz), their amplitude ($E_{pl} \sim 10\text{-}20$ kV/cm) and angular distribution (see below Sec. 3 for a discussion in more detail). Therefore, we can compare the value of $E_{pl}$ taken from [18,19] with that needed to sustain visible (presumably, vacuum) channels in filaments in tokamak T-10.

Figures 1-3 show the results of solving numerically the Poisson equation [23] for arbitrary values of ratio ($l_E/r_D$). It is seen (Fig. 1) that for $U_0 \sim 30$ kV and plasma parameters typical for the periphery of the tokamak, the effective electron density falls down at r ~ 2 mm by seven orders of magnitude, with respect to its background value, and practically disappears at slightly smaller radii.

Thus, the reconstruction of effective voltage bias $U_0$ in the cable from measurements of HF electric field amplitude in plasma of tokamak T-10 [18,19] and Z-pinch [21] shows correlation of the widths (~ 1 mm, in tokamaks, and 0.1 mm, in Z-pinch) of visible straight blocks (presumably cables) with the values of diameters of vacuum channels (around skeleton) which are calculated from quasi-hydrodynamics of a plasma in a HF electric field.

Note also, that the frequency of TEM wave in tokamak T-10 used in calculating the profile of Fig. 1 agrees well with the measurements [20] of the HF electric field spectra outside plasma column (they revealed a permanently present, distinct bump at $\nu_C \sim$ (4-5) $10^9$ Hz, of the width ~ 2 $10^9$ Hz, which always exists in ohmic heating regimes and increases with electron cyclotron heating; this gives $L_{Cab} \approx 3$ cm that is in reasonable agreement with the visible light data from T-10).

## 3. SPECTRAL LINE STARK BROADENING IN PLASMAS AND DIAGNOSTICS OF «WILD CABLES»

In this section we'll consider in more details the setting of theoretical problem of calculation the emission profile of deuterium atoms under conditions that are characteristic for peripheral tokamak and Z-pinch plasmas. Also we discuss physical models that allow to reconstruct the values and frequencies of electric microfields that could be responsible for observational spectra. This allows to reveal the correlation with corresponding values of parameters expected from the «wild cable» concept. From the very beginning we have to assume the possibility of existing of oscillating plasma electric fields acting on the radiator and being the characteristic feature of «wild cables» concept [13].

The Hamiltonian of a hydrogen-like radiator experiencing dipole interaction with high-frequency, **E(t)**, and low frequency, **F**, electric fields and stationary almost uniform toroidal magnetic field **B** has in general the form [24] - [28]

$$H = H_0 - \vec{d}[\vec{E}(t) + \vec{F} + \vec{F}_M(t)] - \vec{\mu}\vec{B}, \qquad (4)$$

where $H_0$ is Hamiltonian of unperturbed radiator, **d, μ** are its electric and magnetic dipole operators accordingly, $F_M(t)$ is the motional electric field $\frac{e}{c}\vec{v}\times\vec{B}$ (e is the electron charge, c is the velocity of light) due to finite thermal velocity **v** of the radiator in the lab frame. It is assumed for simplicity that the second order terms with respect to electric and magnetic fields are neglected in the above equation.

It is known that in plasma the total electric microfield is formed by contributions of so called individual component, which is due to Coulomb electric fields of plasma particles, and collective component which is due to collective plasma electric fields oscillations that are characterized by plasma dielectric function [28]. For low density plasmas in tokamakas and Z-pinch periphery, the contribution of individual component is practically reduced to binary collisional broadening of levels by ions and electrons. Therefore, the contribution from collective microfield component appears to be of relatively more importance making it of special interest in the context of wild cables concept [13].

Under conventional simplifying assumption of density matrix diagonality the polarized profiles of spectral line $I_{\pi,\sigma}(\omega)$ formed in quantum transition from the upper level multiplet of Stark sublevels $\{i\} \in n_{upper}$ (with n being the principal quantum number) to the lower level multiplet of Stark sublevels $\{j\} \in n_{low}$ with polarization $e_{\pi,\sigma}$ are as follows:

$$I_{\pi,\sigma}(\omega) = \frac{1}{\pi} \text{Re} \sum_{i,j} \int_0^\infty dt \cdot \left\langle \exp[-i\omega \cdot t] \cdot <i|\vec{e}_{\pi,\sigma}\vec{d}(0)|j> \cdot <j|\vec{e}_{\pi,\sigma}\vec{d}(t)|i> \right\rangle_{Av} \quad (5)$$

where symbol $<\ldots>_{Av.}$ means the averaging over the large canonical ensemble (in which averages over thermal velocities of perturbing and radiating particles are included as well), $e_{\pi,\sigma}$ is the polarization vector of radiation parallel or perpendicular to the characteristic direction of the system, $\omega$ is the radiation circular frequency. It is important from physical and methodical point of view that the quantization and polarizations are defined in the lab frame.

As is known, there is no general solution of such a problem for arbitrary values and relative directions of electric and magnetic fields. The problem was solved only for some particular model cases that are of practical interest [27]. Below for clarity we neglect the influence of motional Stark and Zeeman effects, because experimental observations to be analyzed seem to be not influenced by these effects.

The spectra formed under action of regular oscillations in the system have as a rule very complicated structure that is especially difficult to interpret. In this case the physically transparent solution is available only in terms of quasi-energy states. This means that in spectral diagnostics one has to use a variety of observational methods to verify by cross-checks the appropriate explanation. It is the complexity of simultaneous action of several mechanisms equally influencing the profile of radiation that make it conventional [25], starting from earlier papers on astrophysical measurements, to consider the profiles of radiation of a fixed polarization and the results of subtracting them from each other to give the so called differential polarization profile $D(\omega)$:

$$D(\omega) = I_\sigma(\omega) - I_\pi(\omega), \quad (6)$$

in order to retrieve and identify the existence of various types of anisotropic oscillating and non-oscillating plasma fields [24-28].

It was natural to define $\pi$ and $\sigma$ polarization with respect to the direction of magnetic field, that is almost constant in peripheral tokamak plasmas under analysis. There are interesting experimental results [18,19] on the study of $I_{\pi,\sigma}(\omega)$ profiles of Balmer series of deuterium in the peripheral tokamak plasmas. The observed spectral lines $D_{\alpha,\gamma}$ showed that $\pi$ and $\sigma$ polarization profiles are different. For example, the spectral line shapes for magnetic field B = 1.65 T and $\pi$ polarization are smooth, while for $\sigma$ polarization they have peaks with separation that exceeds the possible Zeeman splitting [18,19]. For $D_\beta$ profiles, spectra for both polarizations have a smooth, almost similar behavior [18,19].

One of the probable models that could give consistent and non contradictory explanation of those observations is based on assumptions of coexistence of a strong and rapidly oscillating regular plasma field $e_z E_0 \cos \omega t$ and almost quasistatic field $e_x F$ that changes in time much slower [19,27]. If the high frequency $\omega$ is larger than precession frequency of atomic dipole moment in the quasistationary field $\Omega_F \sim (3/2) n^2 F (e a_0/\hbar)$ (where $a_0$ is the Bohr radius) or the oscillating field amplitude $E_0$ is large enough so that $[(E_0 n e a_0/\hbar)\omega]^{1/2}$ is larger than the precession in the field $F$ (see [27]), then it is reasonable to construct the zero-order wave functions of the problem from the wave functions of quasi-energetic states [27]. This approach is characterized by secular equations averaged over time during the period *of 2π/ω*. This averaging automatically leads to the effective field in the form [27]

$$\boldsymbol{F}_{eff.} = J_0(3\,\beta\,n/2)\,F\,\boldsymbol{e}_x, \qquad (7)$$

where $J_0(z)$ is Bessel function of zero order and $\beta = E_0\,ea_0/\hbar\omega$. Thus, in the case when the argument of Bessel function coincides with the location of its zero, the effective splitting due to quasistatic field *F* is completely suppressed by strongly oscillating perpendicular electric field $E_0$ and may serve to determine the ratio $E_0/\omega$. On the other hand, the conditions considered correspond to dominating influence of regular oscillating electric field on the radiator, thus justifying the application of the quasi-energy approach [19]. From the assumption that the frequency $\omega \approx \omega_{ce}$, where $\omega_{ce}$ is electron cyclotron frequency, at field value *B = 1.65 T*, it follows from the value of the first zero of $J_0(z)$ in the case of $D_\beta$ that $E_0 \approx$ *14 kV*. Using this value it is possible then to find the effective fields for $D_{\alpha,\gamma}$ polarization profiles and estimate the *F* value basing on the splitting observed in the $D_{\alpha,\gamma}$ profiles with **π** polarization, that gives value around *F ≈ 20 kV*.

This fact enables to explain the abnormal difference in the behavior of $D_{\alpha,\gamma}$ in comparison with $D_\beta$ profiles, demonstrating the complexity of development of plasma oscillations in tokamak periphery [18,19]. The values of electric fields that match these observations provide the Stark splitting strongly exceeding the Doppler width and Zeeman splitting [18,19] thus justifying the application of a strongly simplified approach. By the way, these measurements and there model interpretation span the new spectroscopic effect consisting in the effective screening, by rapidly oscillating electric field, of the dipole interaction of the radiator with quasistationary electric field *F* [19].

Similar situation in the sense of generation of plasma oscillations was observed under study of some Z-pinch regimes where the spectral profiles observations of peripheral plasmas may be again interpreted in terms of strong perturbations of radiator by regular electric fields of plasma oscillations [21]. The study of $D_{\alpha,\beta,\gamma}$ polarization profiles revealed, together with strong dips in the line body, the difference in the values of half-maximum half-width for the profiles with different polarization [21]. On the other hand, the mean width value was larger than it could follow from the assumption that the main profile is formed due to the action of net quasistatic electric field of plasma ions screened by electrons (individual plasma microfield component) [27]. So, in distinction from the previous case, zero order wave functions of the problem in fact could be formed by this quasistatic stochastic anisotropic microfield $\vec{F}$ originated from low frequency plasma oscillations with anisotropic angular distribution. Hence, in the first approximation the distribution function of quasistatic microfield is approximately described by the Rayleigh distribution of stochastic vector [27].

$$W(F) = \left(\frac{6}{\pi}\right)^{1/2} \frac{3F^2}{F_0^2} \exp\left(-\frac{3F^2}{F_0^2}\right), \qquad (8)$$

under condition that $F_0 \geq 10\,e\,N^{2/3}$, where $F_0$ is the mean squared electric field of low frequency plasma oscillations, and *N* is plasma particle density. The mean level of low frequency plasma oscillations is rather high (~ *10^{-2}*) with respect to the density of thermal energy corresponding to heavy particles $NT_i$ (in those conditions $T_i \geq T_e$), where $T_e$ is the electron temperature and $T_i$ is the ion temperature measured by laser scattering (it lies in the range about *30 eV* [21]). Assuming the distribution of such total low frequency plasma microfield *F* to be known, the whole profile is formed by ion and electron impact collisional broadening of Stark sublevels at a fixed value of *F*. Weighing this picture with the help of microfield probability distribution function $W(\vec{F})$ gives the observable spectral line shape without contribution of high frequency regular oscillating electric field $\vec{E}_0$.

The latter field, however, strongly interacts with the system of Stark sublevels in the vicinity of the following resonances (see [21]):

$$\frac{3}{2} \cdot ea_0 F_{up} n_{up} = \hbar k_{up} \omega, \text{ or } \frac{3}{2} \cdot ea_0 F_{ulow} n_{low} = \hbar k_{low} \omega, \qquad (9)$$

where $F_{up,low}$ are values of quasistatic electric microfield that match the resonance conditions in the multiplet of upper or lower level involved in transition accordingly, while $k_{up,low}$ designates the order of resonance in the upper or lower multiplet of Stark sublevels. The positions of these resonances in the profile being located at a frequency detuning determined by the equation (see [21])

$$\frac{3}{2} \cdot ea_0 F_{up,low} \cdot [n(n_1 - n_2) - n'(n'_1 - n'_2)] = \hbar \cdot \Delta\omega, \qquad (10)$$

where the left side describes the Stark shift of the Stark line component with initial parabolic quantum numbers $\{n,n_1,n_2\}$ and final parabolic quantum numbers $\{n',n'_1,n'_2\}$ at the electric field value $F_{up}$ or $F_{low}$ that satisfy the resonance conditions in the upper or lower multiplet accordingly, $\Delta\omega=\omega-\omega_0$ is the circular frequency detuning from the line center.

Assuming that resonance in the first order $k_{up,low} =1$ is the most intense, it was possible to identify strong dips in the observed profiles of three deuterium Balmer lines emitted from Z-pinch plasmas with positions of resonances corresponding to electron plasma frequency $\omega_{pe}$ [21]. It is very important that these measurements were performed versus time of pinch evolution [21]. Further examination of experimental profiles revealed some kind of the fine structure in the dip's form. For example, the depth of resonance was so large that the intensity of the profile was much lower in this point than in the conventional case, and this structure gave an idea that the value of $E_0$ can not be considered to be negligible. Indeed, dipole interaction of the adjacent Stark sublevels in the field $\vec{E}_0$ leads to an additional repulsion of levels [21], [27]. This results in the appearance of quasienergy splitting of each Stark sublevel. Therefore, there was the opportunity to determine the value of electric field $E_0$ from the measured value of splitting (between two nearest hills surrounding the resonances) which in the assumed model were proportional to $E_0$ [21,27]. The values of electric fields that were retrieved on the basis of those diagnostic methods and models were as follows: *$E_0 \approx 50 \div 80$ kV/cm,* that exceeded the effective thermal equilibrium value, *≈3 kV/cm* (that follows from adiabatic impact broadening by plasma ions), almost by an order of magnitude, *$F_0 \approx 50$ kV/cm*, with the limitation on the frequency range *$\Omega << 3,6\ 10^{12}\ s^{-1}$*, which follows from the quasistaticity condition.

Thus, the used models allow to give an explanation of, at first glance, a strange behavior of polarization profiles simultaneously for three observed lines and to determine the values of anisotropic electric fields. The frequencies of those electric fields could be determined by spectral diagnostics either directly [21] or indirectly [18-19] by means of other diagnostics. For example, in the experiments with Z-pinch plasmas the verification of oscillating electric fields frequencies was possible directly by position of resonances [21].

## 4. CONCLUSIONS

The above analysis shows the Stark diagnostics of hydrogen spectral lines in tokamak and Z-pinch plasmas to be an efficient tool for determining not only the electric field plasma oscillations in a wide range of frequencies (centimeter and millimeter range of vacuum wavelengths) but also for determining the parameters of localized sources of intense HF waves associated with observed long-lived filaments in plasmas. The latter pertains to contribution of the so called wild cables. The «wild cables» concept [13] suggested the observed long-lived skeletal structures in tokamaks and Z-pinches to be caused by a microdust-assembled skeleton. The results (Figs. 1-3) of calculating the widths of vacuum channels produced by the pressure of on-skeleton high-frequency (HF) electromagnetic waves of the TEM type, which thus might protect the skeletons from ambient high-temperature plasma, show that the values of amplitude of such wave appear to be compatible with the measurements of those in-plasma HF electric fields in tokamak T-10 and a gaseous Z-pinch which give observable Stark broadening of hydrogen spectral lines. Thus, the Stark diagnostics of hydrogen spectral lines in tokamak and Z-pinch plasmas gives an opportunity to probe indirectly the parameters of nonlocal (non-diffusion) component of energy transport which is of substantial importance for the problem of energy confinement in fusion plasma facilities.

ACKNOWLEDGMENTS.
The present research is supported by the Russian Foundation for Basic Research (grant 00-02-16453).

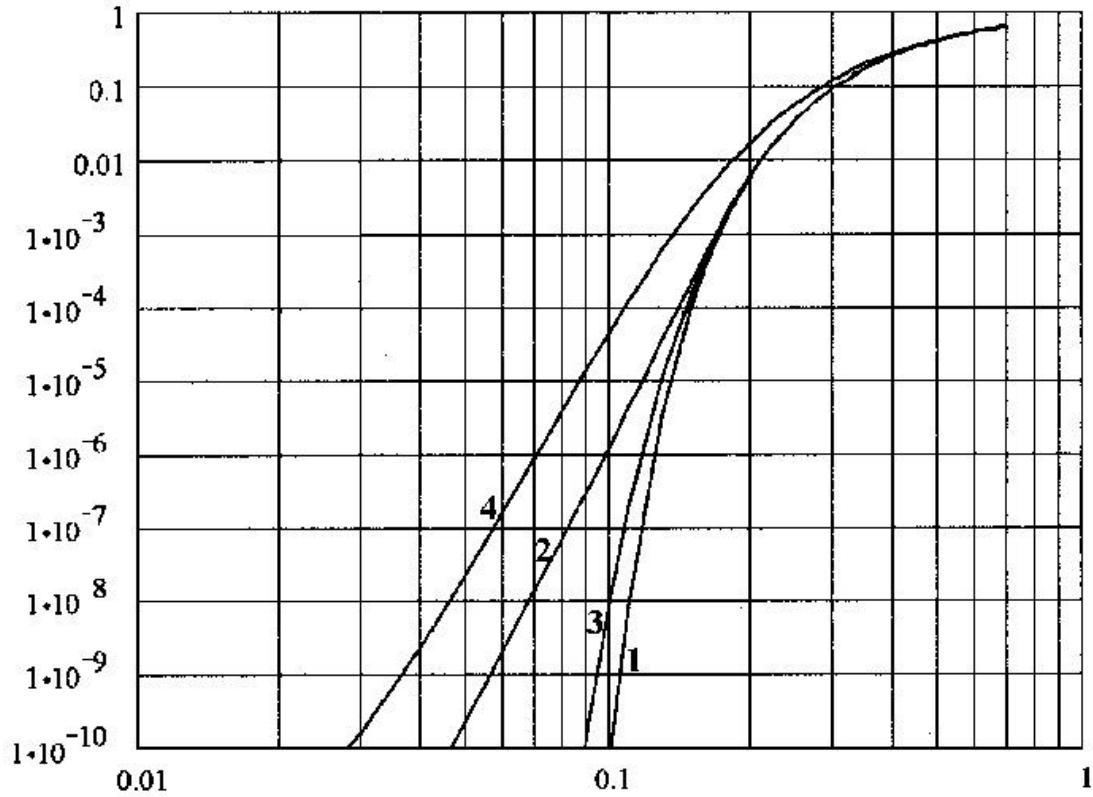

Fig. 1. The profiles of electron (curve 1) and ion (curve 2) particle densities as a function of radial coordinate in a cylindrical circular «wild cable», as calculated from the Poisson equation for a quasi-hydrodynamics [23] of a plasma in a HF electric field, with neglected value of radius of electron HF oscillations ($R_{osc} = 0$). The values of radius are taken in the units $L_{cab}$, where $L_{cab}$ is cable's length, see Eq. (1), the densities are taken in the units of background density. The curve 3 is the plasma density in the limit of plasma quasi-neutrality, and the curve 4 is the curve 1 shifted inward by the local value of $R_{osc}$. The profiles are given for the case of parameters typical for peripheral tokamak plasma, namely background plasma parameters $N_e(0) = 10^{13}$ cm$^{-3}$, $T_e = 100$ eV, and the length of typical straight sections of observed skeletal structures (i.e. hypothetical cable's length) $L_{cab} = 3$ cm. The effective voltage bias in the wild cable is $U = 30$ kV.

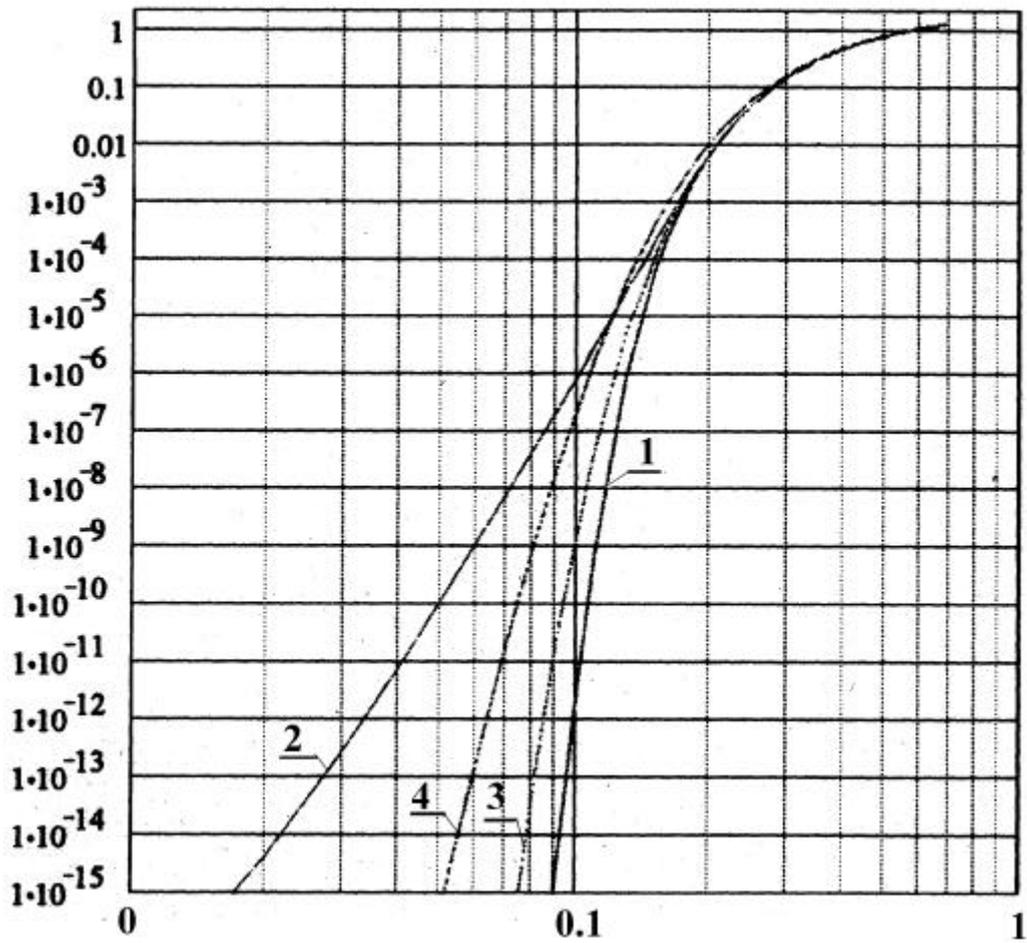

Fig. 2. Radial profiles of particle densities, similar to Figure 1, for the case of parameters typical for the far periphery of a gaseous Z-pinch, namely background plasma parameters $N_e(0) = 10^{15}$ cm$^{-3}$, $T_e = 10$ eV, and the length of typical straight sections of observed skeletal structures (i.e. hypothetical cable's length) $L_{cab} = 1$ mm. The effective voltage bias in the wild cable is $U = 10$ kV.

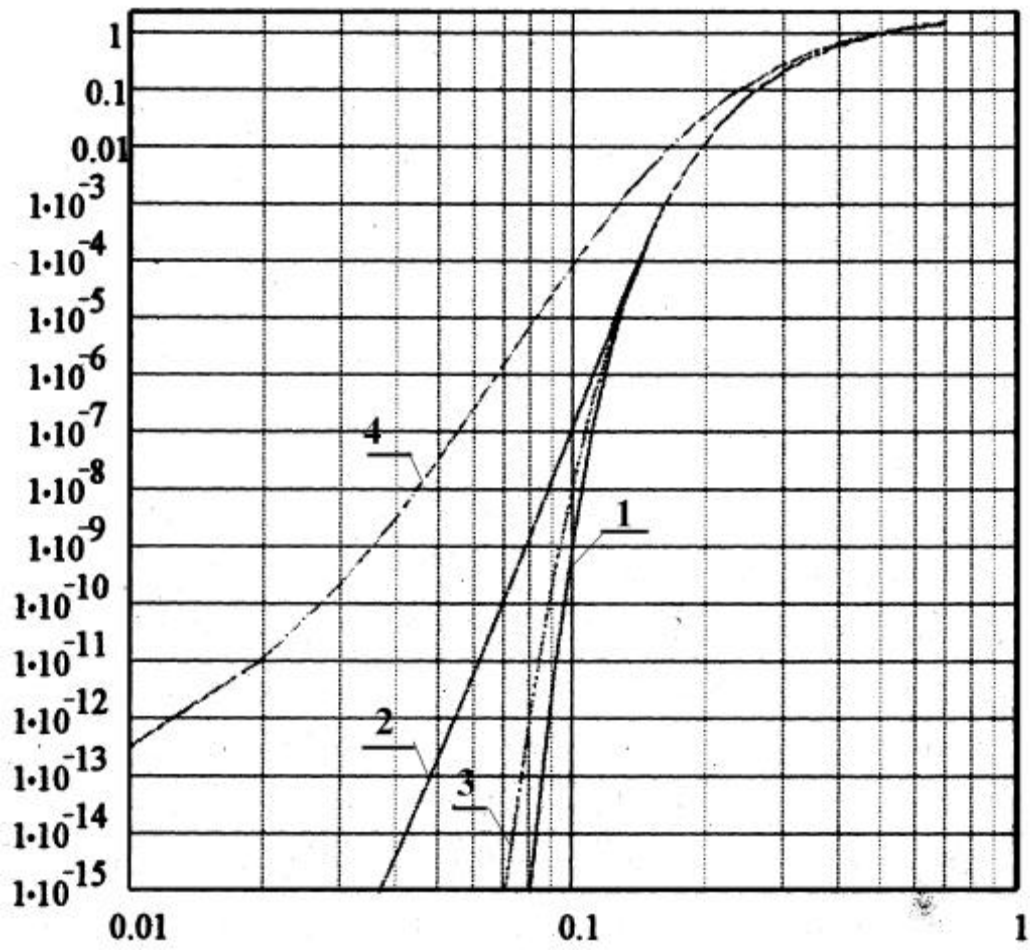

Fig. 3. Radial profiles of particle densities, similar to Figure 2, for the case of background plasma parameters $N_e(0) = 10^{18}$ cm$^{-3}$, $T_e = 100$ eV, which are typical for the core of a high-current gaseous Z-pinch. The effective voltage bias in the wild cable is $U = 30$ kV.